# The $CO_2$ footprint of club soccer in the US.

Aarya Upadhyay, Salem High School, aurium197@gmail.com

## Abstract

I played youth soccer for 11 years starting age at 5. As I progressed through the age groups and got better at the game, I moved up in club levels under the US Club Soccer system. It was during my time as an ECNL player that I began to realize the magnitude of the travel burden associated with local, regional and out of state games. Most of the travel was by road and team members often travelled with minimal car-pooling. Just about that time I was also becoming aware of (and interested in) Climate Change and the various factors impacting climate change including Greenhouse Gases. It was during these travels that I began to track fuel consumption and the idea of estimating the $CO_2$ impact of these soccer road trips was born. In this paper I report my findings through simple calculations and provide extrapolations through simulated scenario experiments. Based on my findings I discuss a few potential solutions that may be considered to reduce the $CO_2$ footprint associated with US youth travel soccer. The idea behind this paper is to raise awareness amongst soccer families and to engage my fellow soccer peers into making the sport we all love as green as possible.

## Introduction

### Youth Soccer in the US is a large community.

Youth Soccer in the USA is a large and complex undertaking. According to one statistic [1,2,3] over 3 million youth players between the ages 5-19 years participate in club level travel soccer. Youth soccer in the US is organized along soccer federations under United States Soccer Federation (USSF). Youth soccer programs such as US Youth Soccer (USYS), US Club Soccer and American Youth soccer Organization (AYSO) are some of the leading youth soccer programs in the country. In addition, most high schools offer varsity soccer programs. Almost all programs, except high schools, run year-round soccer seasons switching between indoor and outdoor leagues and practices depending on the local weather. All Youth Soccer programs have separate girls and a boy's divisions. The soccer league system is multi-tiered. In general, at the competitive level there are several leagues, 4 of the important ones are MLS Next (Academy), ECNL (Elite Clubs National League), USYS and EDP (Elite development Program). Federations run leagues at a National, Regional, and Local level. USYS for example has 13 conferences with ~8 teams per conference. EDP has 3500 teams. MLS Next has ~ 115 clubs with ~490 teams from 6 age groups. ECNL has ~ 80 participating clubs. Clubs organize their teams into age groups with each age group further subdivided into a few subdivisions. For example, an arbitrary age group "0Y" is for players born in the year 200Y and may have teams participating at different accomplishment levels such as MLS Next, ECNL, A-team, B-team State etc. Hence, given that a team typically will have a roster of 18 players, competitive soccer is a very large community.

### Competitive soccer is a commitment to travel.

It is common for teams playing competitive soccer to have practices 3 to 4 times a week with league games scheduled for the weekends. This implies that between practices and games soccer players drive many soccer-miles. Given the varying work schedules of soccer parents as well as their relative geographic residential locations, carpooling is very often not a viable option and soccer players are driven individually to practices and games. Practices are typically held within a 25-mile radius of most players' residences, however, given the competitive nature of the sport and selection into clubs, some players drive over 60 miles for every practice!  Out of state(regional) games are also typically individual player drives since in most cases families like to travel for competitive league games and support their player and team.  In the more competitive levels such as MLS Next and ECNL interstate travel is quite common, some as far as 600 miles away.  Most families choose to drive if distances are within 300 miles, often the cost to fly a family can also become a factor in electing to drive.  In this report [4] the travel commitment of US soccer star Clint Dempsey is mentioned "Starting in fifth grade, Dempsey took six-hour roundtrip journeys to Dallas where his play for elite club Dallas Texans paved the path to stardom". Given all these scenarios, road transportation for soccer is an integral part of the overall US youth soccer experience. It isn't hard to imagine that the cumulative number of soccer-miles driven in the pursuit of the "beautiful game" in the US is not trivial. Additionally, the choice of vehicles used for this travel is very often not the most fuel efficient simply because very often families may not have too many options to when it comes to vehicle choices. It is quite common to see large pick-up trucks and SUVs at practices and games with very often just two riders, the driver, and the player. Another observation is that it is not uncommon to see parked vehicles idling during the entirety of the practice or game as parents of chaperones may prefer to sit in their vehicles to either catch up on work, reading etc. It is also quite uncommon to see EV chargers in the vicinity of practice or game venues.

The rest of this paper is organized as follows. in Section-1 we begin with a brief overview of greenhouse gases and their impacts on global warming and climate change.  In Section-2 we present our analyses of the $CO_2$ footprint from travel soccer for a small population representing a local club.  Next, we use the local statistical models developed in Section-2 to extrapolate over state and then for the entire US. In Section 4, we consider what if scenario simulations and look for potential solutions. We end with a discussion and make the proposal of an ongoing open project inviting all soccer players and high school data analysts to join and contribute.

## 1. Overview of $CO_2$ emissions and contribution from road vehicles

$CO_2$ is a Greenhouse Gas (GhG).  Greenhouse gases trap heat in the atmosphere acting gas an atmospheric blanket.  There are several types of GhG's such as $CO_2$, $CH_4$, $N_2O$, and Fluorinated gases [5]. Each GhG has a Global Warming Potential (GWP) and is calculated relative to $CO_2$.  $CO_2$ is the primary GhG emitted from human activities. In 2021 a total of 37.1 billion metric tons of $CO_2$ were emitted globally [5].  The US is the second largest emitter of $CO_2$ and accounts for approximately 12.6% of global $CO_2$ emissions in 2021 [6]. Based on an EPA report [5], in 2021 $CO_2$ accounted for 79% of all US based GhG emitted from human activities and 35% of this came from the transportation sector. A typical passenger vehicle in the US emits about 4.6 metric tons of $CO_2$ per year.  This assumes an average fuel economy of 22.2 miles per gallon (mpg) and 15,000 miles driven per year [7], however, this number appears to vary depending on the organization reporting this statistic. There are some excellent resources to understand the relationship between $CO_2$ and GhG in general and global warming [8, 9].  It is generally accepted that a global warming to 1.5°C above preindustrial levels and exceeding this limit would trigger a series of irreversible changes to the global climate system. This 1.5°C limit is also referred to as eth tipping point.  Based on reports by World Meteorological organization (WMO) global temperatures are on track to exceed this threshold and will most likely be between 2.4°C and 2.6°C.  In order to prevent this breach of the 1.5°C   limit the International Panel on Climate Change (IPCC) recommends that a reduction in GhG emissions by approximately 50% by 2030 may be necessary. Given this urgency all and any efforts to reduce GhG must be adopted, including contributions from club soccer and other similar activities.

## 2. An assessment of $CO_2$ emissions from club/travel soccer

In this section I present a detailed assessment of the CO2 footprint from US Club soccer. As I discussed in the introduction, Club Soccer in the US is a massive undertaking with varying travel networks. Hence, in order to perform a systematic and accurate assessment of youth soccer related $CO_2$ it is important to have access to relevant data. In this Section we discuss the data used for this study and also discuss the statistical methods used for making projections.

### Data

In order to perform an analytical assessment, I looked for soccer travel related data resources. As it turns out clubs and teams do not post detailed games and practice schedules on their websites, hence collecting detailed travel information is not straightforward. Most teams use utilities like "TeamSnap" [10] to track practice a game events and these are open only to the team. In preparation for this report, I had saved my travel information for 1 season when I played in the ECNL league for a club. For the purposes of anonymization, I refer to the players as $P_i$ with $i = (1 \ldots M)$. The data set includes destinations common destination, $D_k, k = (1 \ldots N)$ for all practices as well as matches during one season that is approximately 11 months long. This gives us about $N = 192$ unique travel events for each player.

The strategy for calculating the $CO_2$ footprint is simple as shown in the pseudo code steps below:

- ❖ **def Soccer_CO2_Calculator ():**
- for k in Num of unique games (N)
  - for i in Num of players (M)
    - get Origin $O_i$ for each player
    - get destination $D_k$ for reach game event
    - get distance travelled, $S_{i,k}$ [miles] between $(O_i \rightarrow D_k \rightarrow O_i)$ for each player for each game
    - get fuel consumption for each player for each game $F_{i,k}$ [miles/gal] → use vehicle $V_i$ used by each player and the fuel economy $FE_i$ associated with the vehicle $V_i$. Fuel consumption is then calculated as $F_{i,k} = \frac{S_{i,k}}{FE_i}$ [gallons]
    - get $CO_2$ impact for each player for each game by converting fuel consumption to $CO_2$ as → $m_{i,k}^{CO2}[gm] = F_{i,k}[gal] * 8887 \left[\frac{gm}{gal}\right]$

From the pseudo code above, it is clear that in order to calculate travel distances, the "origin" geo-location, $O_i$, of each player, is needed. The destination, $D_k$ is common for all players and varied between travel events, except for practices. Typically practice related drives are shorter than league game related travel. Origin data for each player is typically only available at the club level and not made public. If exact player level origin data is not available, it is reasonable to assume the club home location as the common origin for all players. This assumes that all players on the team live within a 25-mile "local" radius of the club. However, this may underestimate travel distances for players that live outside the reasonable local radius. In the data set used in the analysis in this report there were a few players that did travel long distances even for practice (similar to the Clint Dempsey example). Distances were calculated for shortest route using any road distance calculator such as Google maps [11]. Note that if using API's the origin-destination locations would have to be converted into latitude-longitude co-ordinates. Figure 1 shows an example of soccer related drives for a single player with destinations and origin indicated on a geographic map. These drives include regular practice as well as match-play. For this player, $P_1$, travel distances ranged between 10 miles to 300 miles.

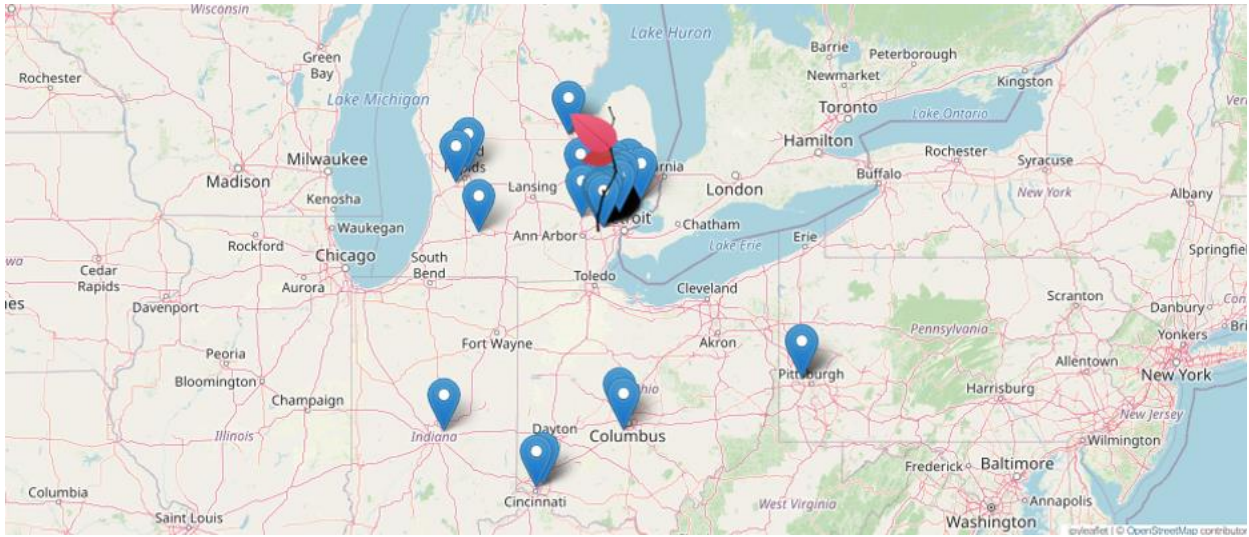

Figure 1: Travel map for a single participant for one season. Destinations are shown in blue; origin is indicated by the Leaf Icon.

Figure 2 shows the round-trip distances travelled (1000 miles) by the entire team over the full season. While most of the team aggregated between 10000 and 15,000 miles over the season there were some players (P14, P18) that were aggregating in excess of 30,000 miles! Note that the average miles driven per year by a typical family in the US is approximately 15,000 miles.

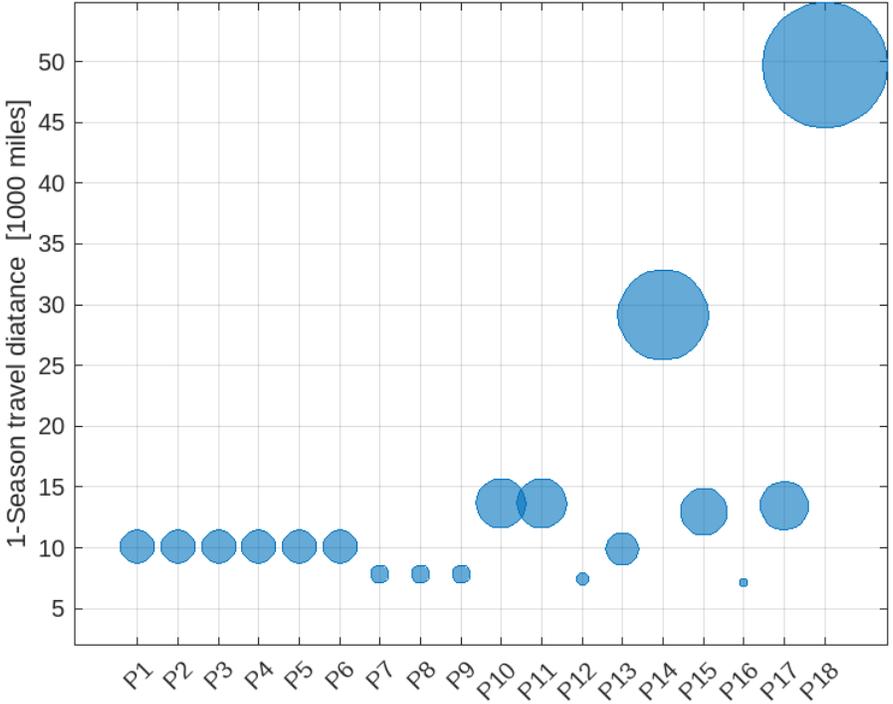

Figure 2: Single season total travel distance (round trip) for all players.

Single player statistics are further evaluated for $P_1$ as a representative (mode) illustration of the player group. Figure 3 shows the one-way travel distances for player $P_1$ as a fraction of all trips during one season. For $P_1$ most of the soccer travel is within the 8-to-60-mile range and is related to practices and in-state games. In contrast $P_{18}$ travels between 120 miles to 165 miles for similar soccer events.

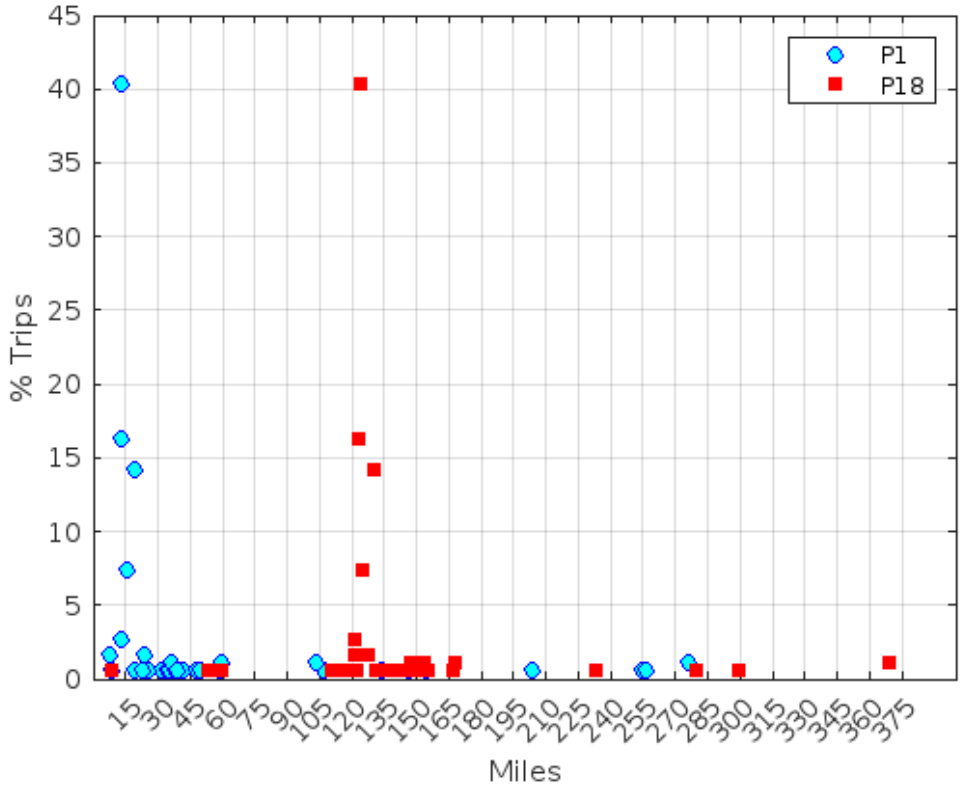

*Figure 3: One-way miles traveled as a fraction of all trips over 1 season for players $P_1$ and $P_{18}$, representing two travel extremes.*

Next, we evaluated the same trends for the entire team. Recall that each player would have a similar travel map except with a different origin. Figure *4* shows a pie-chart representation of all travel and the associated soccer event. Over 55% of all travel for the team was for practice events, 37% of the trips were for in-state competitive games and 8% of the travel was for out-of-state competitive games. We also indicate the total driven miles for players $P_1$ and $P_{18}$ for each of the Pie-Chart sectors. We note that while practices represented 55% of total travel for both players, the implications for total travel for the year were quite different. While practice related travel represented approximately 44% of all travel miles for $P_1$ it represented 77% of all travel miles for $P_{18}$! This is not surprising given that $P_1$ is local to the club while $P_{18}$ is over 100 miles away. $P_{18}$ is a "Dempsey like player. Equivalent travel fractions for the other soccer event sectors are also shown. Interestingly the impact of out of state travel on $P_{18}$ was not as dramatic given the proximity of $P_{18}$ to the out of state event venues.

Since the fuel consumption and hence the $CO_2$ emissions from a vehicle depends on the vehicle type, we used the vehicle class used by the various players to determine these numbers.

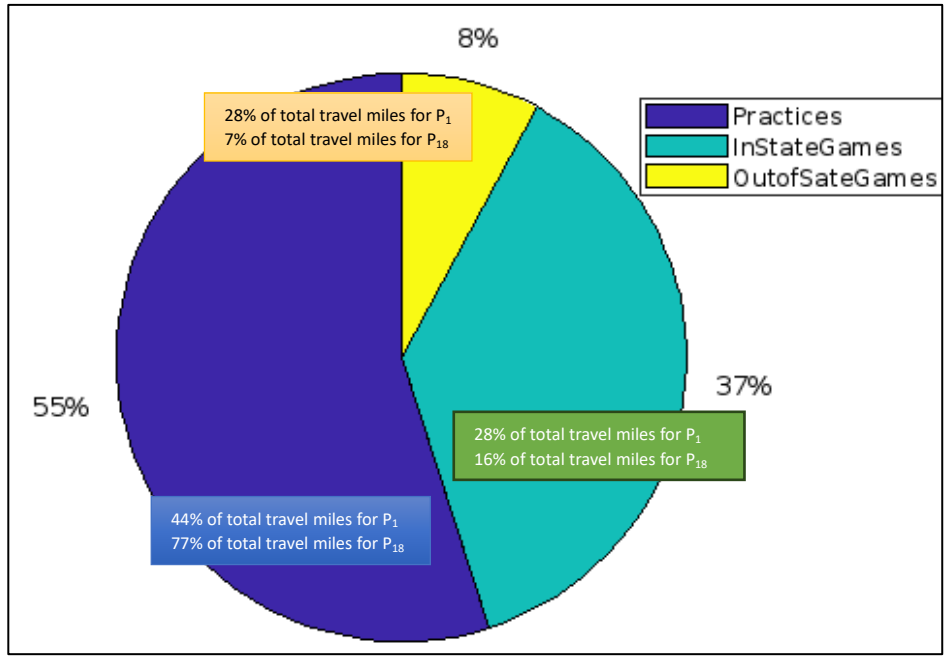

Figure 4: Pie chart for travel by soccer events.

Figure 5 is a pie-chart showing the distribution of the vehicle classes used by the players. It is observed from this group that SUV's and pick-up trucks are the most popular mode of transportation making up 77% of the vehicle usage, while 17% of the players used cars and only 6% of the family's used vans. Notice that there are no Electric or hybrid electric vehicles in the mix since this data is from 2020 when xEV's were not as prevalent as they are today. Table 1: Rated Combined Fuel economy [miles/gallon] for various vehicle types.

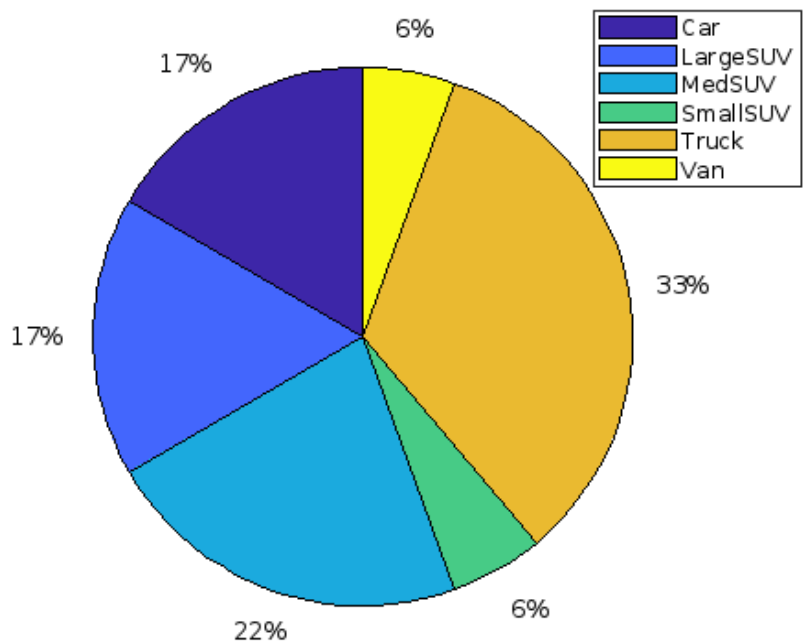

Figure 5: Vehicle preference by players. The type of vehicle impacts Fuel Economy hence $CO_2$.

| Vehicle type | Combined Fuel Economy [mpg] |
|---|---|
| Car (mid-size) | 27 |
| SUV-small | 27 |
| SUV-Medium | 24 |
| Van | 20 |
| SUV-large | 18 |
| Truck | 16 |
| Car-HEV | 104 |
| Car-EV | 130 |

Table 1: Rated Combined Fuel economy [miles/gallon] for various vehicle types.

Table *1*. shows the rated Fuel Economy (FE) in miles/gallon [mpg] for the vehicle classes used by the team. Since vehicles FE ratings [12] are listed for Highway driving and City driving separately we use the combined FE numbers that are also reported. It is also noted that the FE numbers vary across manufacturers and vehicle variants. We arbitrarily pick some middle of the pack numbers (combined) to run our $CO_2$ analyses. For reference we also include the equivalent FE of a Hybrid Electric Vehicle (HEV) and an all-Electric Vehicle (EV).

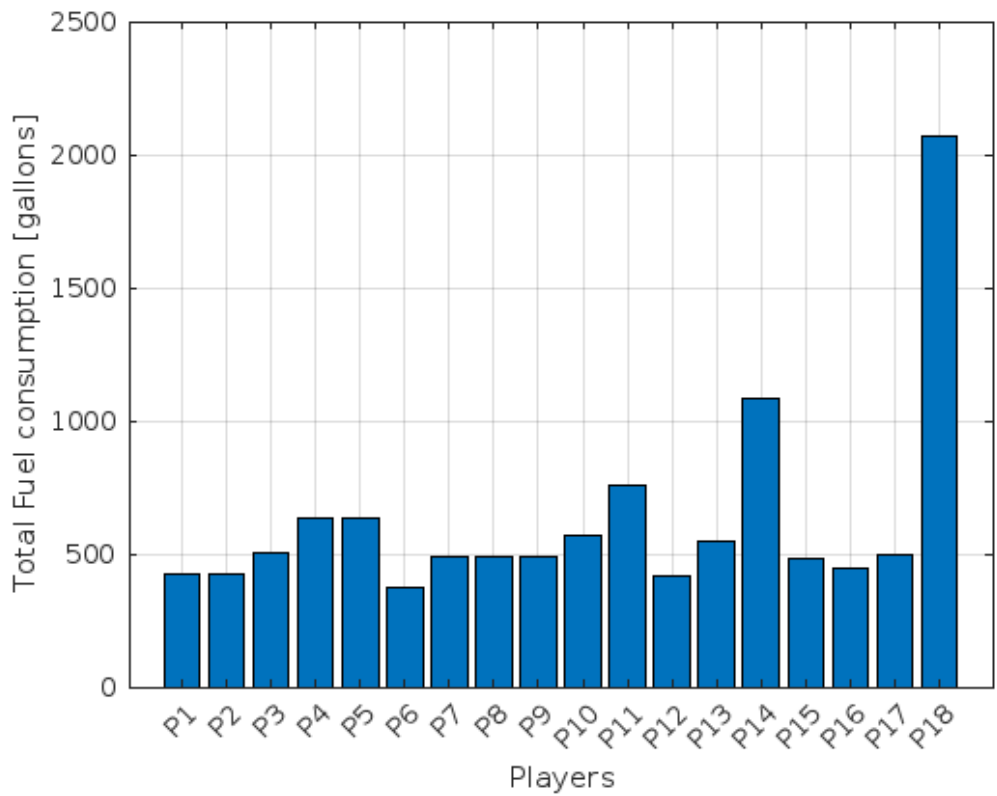

Figure 6: Total annual Soccer travel related Fuel Consumption by individual player.

In Figure 6 we show the total soccer travel related fuel [gallons] consumed by each player. As expected $P_{18}$ has the highest consumption by a large margin. This is a combination of the distance travelled as well as the vehicle type used (Truck). The fuel consumption for each player is converted to a $CO_2$ footprint based on the simple conversion factor of 8887 grams of $CO_2$ per gallon of fuel consumed. We show the

$CO_2$ results in Figure 7. In Figure 8 we show the cumulative $CO_2$ footprint over all players with a total team $CO_2$ footprint of about 100 metric tons annually and player $P_{18}$ has a 18% contribution alone.

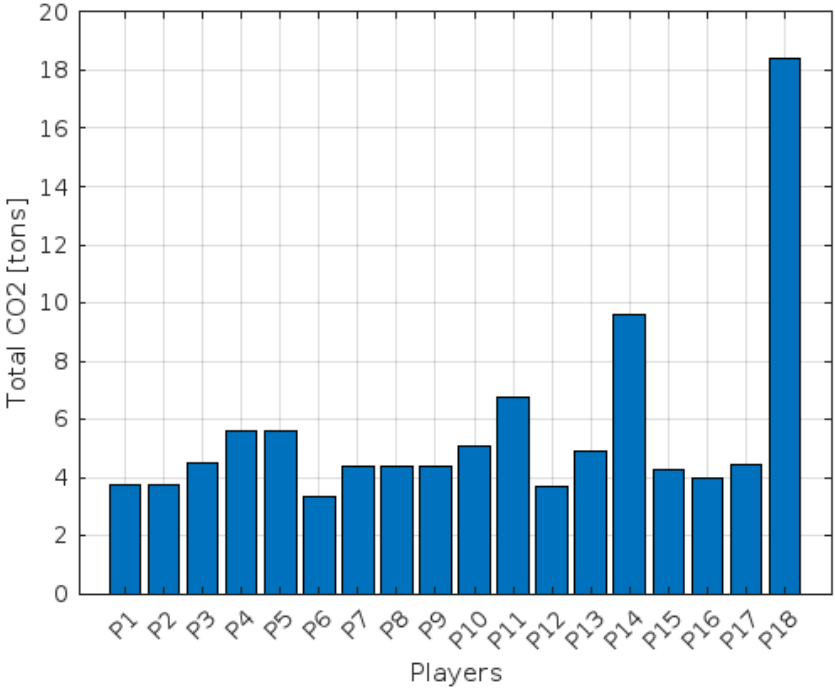

Figure 7: Total annual Soccer travel related $CO_2$ footprint by individual player.

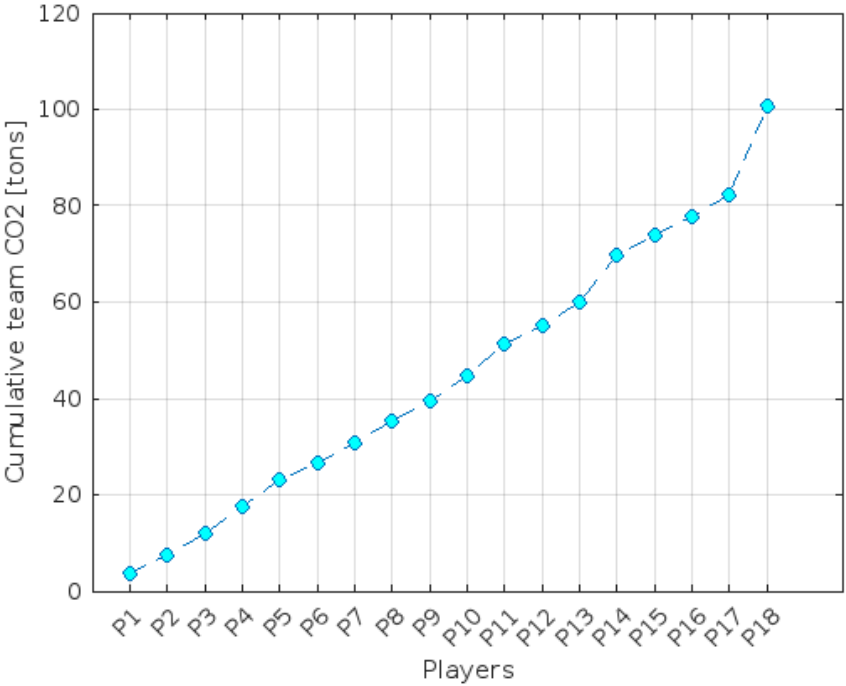

Figure 8: Cumulative sum of annual Soccer travel related $CO_2$ footprint for the entire team.

Compared to global $CO_2$ annual emissions of approximately 7 Giga Tons this may not seem like a large number. However, when we consider all teams across all states, we see a different picture emerge. Since it was difficult to get an exact estimate of soccer teams across the US, we performed a simple study based on our findings for the single club we have reported so far. Using the value of 100 metric tons of $CO_2$ annually as representative we make some extrapolations based on the following assumptions:

- Number of travel clubs per state = 25
- Number of teams by age group = [1:14]. This considers age groups 6 to 19 and form the basic team units for each club.
- Number of team variants per age group = 3. This considers levels like A team, B team etc.
- Number of Soccer playing states = 52.

Based on these assumptions we ran some simple calculations to get a sense of the total $CO_2$ footprint of Club-Soccer in the US. Figure 9 shows the estimated total annual $CO_2$ from soccer related travel in the US as a function of the number of basic team units in each club. Based on this simple calculation we can estimate that the total $CO_2$ footprint of US club soccer may be in the region of 6 mil metric tons per year. This amounts to less than 0.1% of global annual $CO_2$ emissions from road vehicles. These results are shown in Figure 10.

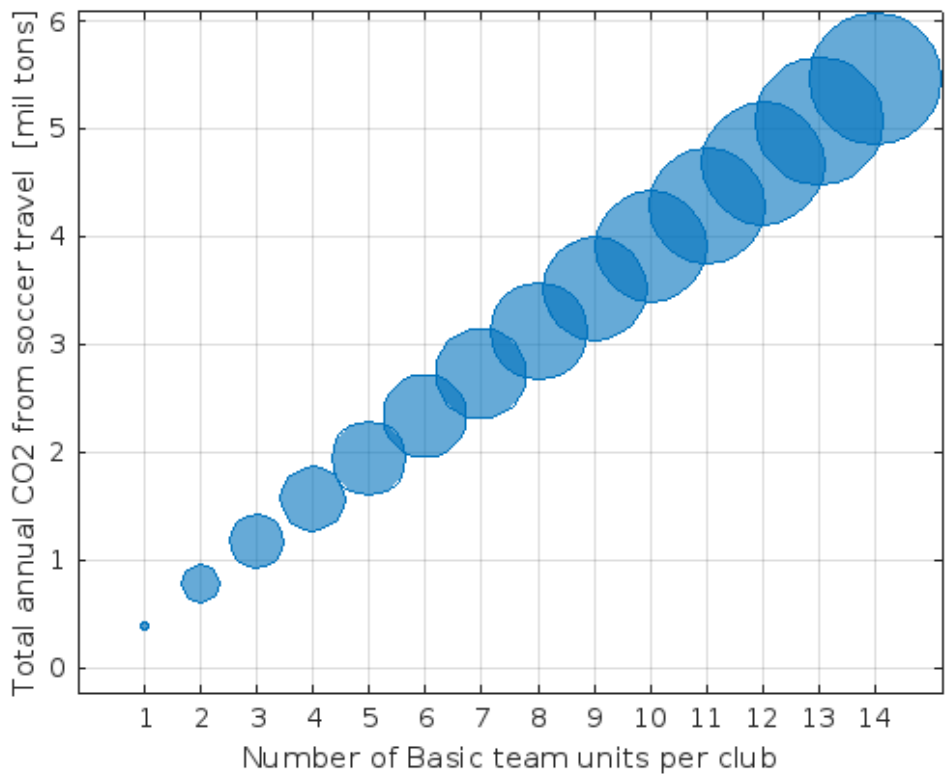

Figure 9: Estimated total annual $CO_2$ [mil metric tons] from US club soccer related travel as a function of number of basic team units per club.

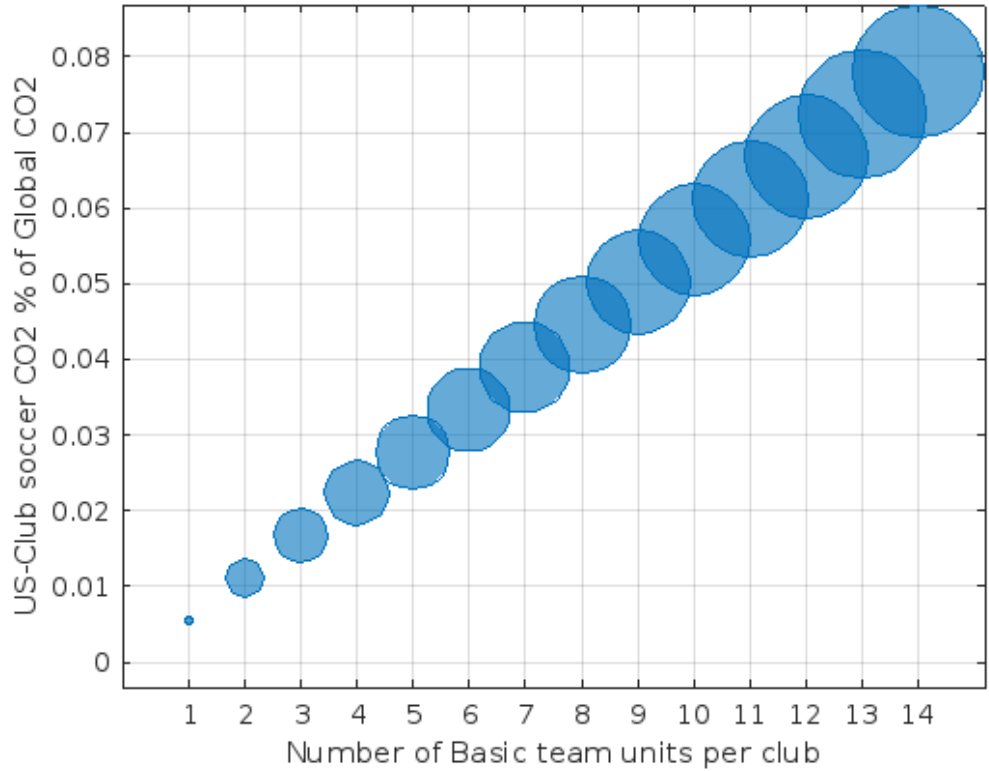

Figure 10: Estimated total annual $CO_2$ from US club soccer related travel as a fraction of global annual vehicular $CO_2$ emissions.

## Conclusions and next steps

This began as a simple exercise to satisfy two of my passions that are analyzing data using modern data science tools and sustainability issues concerning our planet. The idea of using soccer travel data for this analysis allowed me to not only experiment with data analytics but also answer the question around the $CO_2$ footprint from travel soccer.  Granted that the assessments presented are simplistic and several factors are not considered, such as many parents idle their vehicles while waiting for a practice session to complete. This definitely adds to the $CO_2$ emissions but is not considered in this study.  As next steps I plan to consider strategies that will reduce the existing footprint such as carpooling and the use Electric or Hybrid Electric vehicles. While researching the vehicle specific Fuel Economies as reported in [12] it was interesting to learn that the fuel economies for the same class of vehicle varied (often dramatically) between manufacturers.  While strategies such as car-pooling have been around, they are unfortunately not used often. Bus transportation to out-of-state games may also be used but this does not address the issue of parents who would also like to travel with the team.  Strategies that better optimize the travel to minimize travel times via optimal league games organization may also be considered. Team management Apps may offer such services.  I also believe that with EV charging infrastructure at games/practice venues more people will use such vehicles and thereby reduce the $CO_2$ impact.  As I find more time, I hope to share my code and anonymized data and invite the soccer playing community to join in and contribute, hopefully acquiring data resemblant of the country, and possibly even internationally.  One extension of this work may be to consider all soccer including international club soccer and the associated $CO_2$ from flying.  Extensions to other sports such as Tennis, Swimming, American Football, Baseball etc. is also interesting.  For those interested this will be a great learning experience while also helping improve the sustainability of Soccer, the sport we all love.

## Acknowledgements

I would like to thank Ratna Bearavolu and Devesh Upadhyay for their guidance in writing this paper.  I would also like to acknowledge all my soccer teammates and my club for their support.